\documentclass[english]{article}
\usepackage[]{fontenc}
\usepackage[latin1]{inputenc}
\usepackage{amsmath}
\usepackage{graphicx}

\makeatletter

\providecommand{\tabularnewline}{\\}

\newcommand{\lyxaddress}[1]{
\par {\raggedright #1
\vspace{1.4em}
\noindent\par}
}

\usepackage{times}

\usepackage{color}

\usepackage{babel}
\makeatother
\begin{document}

\title{Effective mapping of spin-1 chains onto integrable fermionic models.
A study of string and Néel correlation functions}

\author{C Degli Esposti Boschi$^{\text{a,b}}$, M Di Dio$^{\text{b,c}}$,
G Morandi$^{\text{b,c}}$ and M Roncaglia$^{\text{d}}$}

\maketitle
\vspace{1.4em}

\lyxaddress{$^{\text{a}}$CNR, Unità CNISM di Bologna, viale Berti-Pichat, 6/2,
I-40127, Bologna, Italia}


\lyxaddress{$^{\text{b}}$Dipartimento di Fisica dell'Università di Bologna,
viale Berti-Pichat, 6/2, I-40127, Bologna, Italia}

\lyxaddress{$^{\text{c}}$INFN, Sezione di Bologna, viale Berti-Pichat, 6/2,
I-40127, Bologna, Italia}

\lyxaddress{$^{\text{d}}$Max-Planck-Institut für Quantenoptik, Hans-Kopfermann-Str.
1, D-85748, Garching, Germany}

\begin{abstract}
We derive the dominant contribution to the large-distance decay laws
of correlation functions towards their asymptotic limits for a spin
chain model that exhibits both Haldane and Néel phases in its ground
state phase diagram. The analytic results are obtained by means of
an approximate mapping between a spin-1 anisotropic Hamiltonian onto
a fermionic model of noninteracting Bogoliubov quasiparticles related
in turn (via Jordan-Wigner transformation) to the XY spin-1/2 chain
in a transverse field. This approach allows us to express the spin-1
string operators in terms of fermionic operators so that the dominant
contribution to the string correlators at large distances can be computed
using the technique of Toeplitz determinants. As expected, we find
long-range string order both in the longitudinal and in the transverse
channel in the Haldane phase, while in the Néel phase only the longitudinal
order survives. In this way, the long-range string order can be explicitly
related to the components of the magnetization of the XY model. Moreover,
apart from the critical line, where the decay is algebraic, we find
that in the gapped phases the decay is governed by an exponential
tail multiplied by power-law factors. As regards the usual two points
correlation functions, we show that the longitudinal one behaves in
a {}``dual'' fashion with respect to the transverse string correlator,
namely both the asymptotic values and the decay laws exchange when
the transition line is crossed. For the transverse spin-spin correlator,
we always find a finite characteristic length which is an unexpected
feature at the critical point.

The results of this analysis prove some conjectures put forward in
the past. We also comment briefly the entanglement features of the
original system versus those of the effective model. The goodness
of the approximation and the analytical predictions are checked versus
density-matrix renormalization group calculations.

\vspace{1.4em}

or topological 
\end{abstract}

\section{Introduction\label{sec:Intro}}

The Haldane phase \cite{kt1992}, found in many low-dimensional spin
systems, has attracted a great amount of attention in the last two
decades both from the theoretical and from the experimental points
of view. Its genuine quantum nature is signalled by two characteristic
features. First, the excitation spectrum above the ground state (GS)
displays a finite energy gap and, second, one can identify suitable
long-ranged string correlation functions that measure a hidden topological
order of the phase. The most intuitive idea to understand the physical
features of the Haldane phase is probably the \textit{spin liquid}
picture \cite{t1991}: In a spin-1 chain with Heisenberg interactions
and quantization axis directed along $z$, let us assign the presence
of an effective spin-1/2 particle with spin pointing up (down) if
at the $i$-th lattice site $S_{i}^{z}=+1$ ($-1$) and no particles
if $S_{i}^{z}=0$. The Haldane phase is then interpreted as a liquid
in which these effective particles carry no positional order along
the chain but still retain antiferromagnetic (AFM) order in their
effective spins. The positional disorder is associated with the absence
of long-range order in the usual spin-1 correlation functions\[
{\cal C}_{\alpha}(R)\equiv(-1)^{R}\langle S_{i}^{\alpha}S_{i+R}^{\alpha}\rangle\;,\;\;\alpha=x,y,z\]
 whereas the spin-1/2 magnetic order that we would get if all the
sites with $S_{i}^{z}=0$ were taken off from the chain is measured
by the asymptotic value of the string correlators \cite{dr1989}:\begin{equation}
{\cal O}_{\alpha}(R)\equiv\langle S_{i}^{\alpha}{\rm e}^{{\rm i}\pi\sum_{j=i+1}^{i+R-1}S_{j}^{\alpha}}S_{i+R}^{\alpha}\rangle\;,\;\;\alpha=x,y,z\label{eq:scf}\end{equation}
 for $R\to\infty$. Interestingly enough, the Haldane gap has been
interpreted as the excitation energy associated with a {}``spinon''
(or kink) with respect to the hidden order \cite{em1994}. The nonvanishing
values of the string-order parameters (SOP)\[
{\cal O}_{\alpha}\equiv\lim_{R\to\infty}{\cal O}_{\alpha}(R)\]
can be understood as a spontaneous breaking of hidden (nonlocal) Z$_{2}$
symmetries of the $\lambda-D$ Hamiltonian, as discussed thoroughly
by Kennedy and Tasaki \cite{kt1992}. From a numerical inspection
of the string correlation functions (\ref{eq:scf}) computed on the
first excited state with $S_{{\rm tot}}^{z}=1$, rather than on the
GS, Elstner and Mikeska \cite{em1994} argued that this excited wave
function is characterized by a transition region with vanishing string
correlations that connects two asymptotic limits with symmetry breaking
and different values of the hidden order. In a field-theoretic approach
to spin-1/2 Heisenberg chain \cite{gnt1998} such a kink is described
as an effective particle - a soliton - moving with relativistic dispersion
relation. When the system is moved away from criticality, due to the
action of a relevant field, the soliton acquires a nonvanishing mass
or an energy gap, in the condensed matter language. In ref. \cite{cvetal2006}
some of the authors have proposed a picture of the states that form
the Haldane triplet at the isotropic point in terms of massive solitons
and their bound states arising in the sine-Gordon formulation, valid
in the neighbourhood of the critical line that marks the limit of
the Haldane phase towards the large-$D$ one (see below). The first
solid numerical evidence of a nonzero Haldane gap has been provided
by White and Huse \cite{wh1993} using the by now celebrated density-matrix
renormalization group (DMRG) method.

Actually, the Haldane phase is not restriced to spin-1 systems and
can be found, for example, in spin-$S$ Heisenberg chains for every
integer value of $S$. According to ref. \cite{qlsc2003} the gap
vanishes as the classical limit $S\to\infty$ is approached as $\Delta\propto S^{-1}\exp(-\pi S)$
while the behaviour of the string order is more subtle: in order to
have a nonzero value one has to generalize the string correlation
function of equation (\ref{eq:scf}) using not $\pi$ in the exponential
but $S$-dependent optimal angles $\theta_{n}=(2n+1)\pi/S$ with $n=0,1,\dots,S-1$.
Again, when $S\to\infty$ the resulting values of ${\cal O}_{\alpha}(\theta_{n})$
tend to zero.

It is interesting to examine also how the features of the Haldane
phase are destroyed by varying the parameters of the Hamiltonian out
of the isotropic spin-$S$ Heisenberg model ($S$ integer). In this
paper we shall stick from now on to the case $S=1$ and consider two
types of anisotropies along $z$: Ising-like interactions (parametrized
by $\lambda$) and single-ion terms (parametrized by $D$)\begin{equation}
H=\sum_{i}\vec{S}_{i}\cdot\vec{S}_{i+1}+(\lambda-1)S_{i}^{z}S_{i+1}^{z}+D(S_{i}^{z})^{2}.\label{eq:HlamD}\end{equation}

\begin{figure}
\includegraphics[clip,width=7cm]{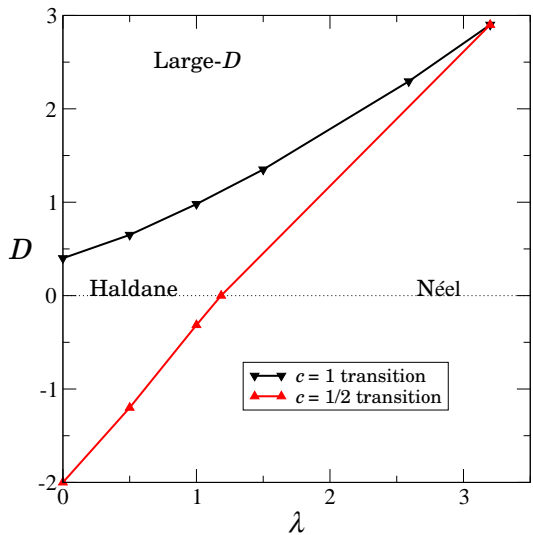}

\caption{Ground-state phase diagram for the model (\ref{eq:HlamD}) in the
AFM region $\lambda\ge0$. The three phases are defined in the text.\label{fig:pd}}
\end{figure}

The phase diagram of this model has been investigated in various papers
with different approaches \cite{chs2003,dr1989,kt1992}. In order
to fix the ideas we will refer to our recent determination \cite{cvetal2006},
reported (in a simplified form) in figure \ref{fig:pd}. Fixing a
nonnegative value for $\lambda$ and varying $D$ we encounter three
gapped phases: the Large-$D$ one in which ${\cal O}_{\alpha}=0\;\forall\alpha$
indicating the absence of magnetic order in the effective spin-1/2
particles. Their positional degrees of freedom are also disordered.
In the Haldane phase the spatial disorder persists but magnetic order
emerges. As a consequence both longitudinal and transverse string
order parameters (SOP) become nonzero: ${\cal O}_{\alpha}\neq0$.
As pointed out in ref. \cite{bgo2004} on the basis of an exact solution
for an integrable variant of (\ref{eq:HlamD}) with $\lambda=0$ and
in-plane anisotropy, the excitation gaps in the Large-$D$ and in
the Haldane phases have a rather different nature. Despite the fact
that they are both found within the sector $S_{{\rm tot}}^{z}=1$,
the former corresponds to a flip of a single spin out of the $xy$
plane while the latter is related to the breaking of a two-site singlet
composing the resonant-valence-bond GS similar to the one of the spin-1
chain exactly solved by Affleck, Kennedy, Lieb and Tasaki \cite{aklt1988}.
Finally, by decreasing further the value of $D$, we pass in the Néel
phase where both positional and magnetic degrees of freedom orders
are signalled by a nonvanishing (spontaneous) magnetization along
$z$\[
M_{z}^{2}\equiv\lim_{R\to\infty}{\cal C}_{z}(R).\]
At the same time ${\cal O}_{z}\neq0$ but ${\cal O}_{x,y}=0$. Den
Nijs and Rommelse (\cite{dr1989}, Sect. IIE) introduced yet another
less-familiar string correlation function without spins at the ends\[
G_{H}(R)\equiv\langle{\rm e}^{{\rm i}\pi\sum_{j=i}^{i+R}S_{j}^{z}}\rangle\]
 and argued that $G_{H}(\infty)=0$ in the Haldane phase but $G_{H}(\infty)\neq0$
in the Large-$D$ and Néel ones.

Hence we may select, equivalently, the pairs $({\cal O}_{z},{\cal O}_{x})$
or $({\cal O}_{z},M_{z})$ as order parameters to classify the three
types of behaviour. The universality classes associated with the two
transition lines will be frequently denoted using the language of
conformal field theory (CFT - see, for instance, \cite{h1999,chs2003,cvetal2006}),
in particular by specifying the central charge $c$. We interpret
the fully-disordered large-$D$ phase with $({\cal O}_{z}=0,{\cal O}_{x}=0)$
and $({\cal O}_{z}=0,M_{z}=0)$ as a spin gas. By crossing the $c=1$
line we enter the Haldane phase where the effective spin-1/2 experience
a first magnetic ordering: $({\cal O}_{z}\neq0,{\cal O}_{x}\neq0)$
and $({\cal O}_{z}\neq0,M_{z}=0)$. Then, loosely speaking, at the
$c=1/2$ line the spin liquid crystallizes and the fully-ordered Néel
phase can be interpreted as a spin solid with $({\cal O}_{z}\neq0,{\cal O}_{x}=0)$
and $({\cal O}_{z}\neq0,M_{z}\neq0)$. Note the interchanged role
of ${\cal O}_{x}$ and $M_{z}$ (see below). In the Néel and Haldane
phases $G_{H}(\infty)$ refers to the positional order of nonzero
spins \cite{dr1989}, so that it vanishes in the Haldane phase but
$G_{H}(\infty)\neq0$ in the Néel one.

In order to determine the SOP numerically one has to extrapolate to
the thermodynamic limit and to infinite distance the data computed
on necessarily finite samples. However, apart from the qualitative
statements made in ref. \cite{dr1989} about the exponential decay
of the string correlation functions (as reported in table \ref{tab:dlscf}),
the available literature contains scarce information about the spatial
behaviour of such correlators and the extrapolation may become problematic,
especially close to the transition lines where the bulk correlation
length becomes very large. In a particular case, namely the transition
from the Large-$D$ to the Haldane phase, the low-energy physics is
described by a compactified free boson field theory ($c=1$ CFT).
Once the compactification radius is known in some other independent
way, one can read off the decay exponent of the string correlation
functions from the set of scaling dimensions of the possible vertex
operators. Interestingly, it turns out \cite{ah1992,debeor2003} that,
even if the lattice model has periodic boundary conditions (PBC),
the vertex operators to be associated with string correlators belong
to the sector with \textit{antiperiodic} boundary conditions.

\begin{table}

\caption{Decay laws of string correlation functions defined in the text, according
to Sect. IIE of ref. \cite{dr1989}. To be compared with the results
of this paper, including the explicit form of the algebraic prefactors,
as reported in table \ref{tab_fitt}. \label{tab:dlscf}}

\begin{tabular}{@{}lll}
\hline 
Phase&
${\cal O}_{z}(R)$&
$G_{H}(R)$\tabularnewline
\hline 
Haldane&
Expon. to $\neq0$&
Expon. to $0$ \tabularnewline
Néel&
Expon. to $\neq0$&
Expon. to $\neq0$\tabularnewline
\hline
\end{tabular}

\end{table}

The main purpose of this paper, instead, is to address the spatial
behaviour of spin-spin and string correlation functions in the Haldane
and Néel phases, making use of a solvable theory of spinless fermions.
Starting from well inside the Néel phase, where the density of sites
with $S_{i}^{z}=0$ is negligible, we approximate the problem by assuming
that the hidden magnetic order is frozen so that a given contribution
to the GS wavefunction can be described by occupation numbers: no
fermions if $S_{i}^{z}=0$ and one fermion when $\vert S_{i}^{z}\vert=1$,
no matter the orientation, which is dictated by the underlying string
order. The details of this approach will be presented in Sect. \ref{sec:Mapping};
actually it is very close to what done by G\'{o}mez-Santos in ref.
\cite{g1989}. The difference here is that we include also the single-ion
anisotropy term and, in fact, the two formulations are related by
a particle-hole transformation. The novelty is that we work out in
detail the mapping of the spin-spin and string correlation functions
(subsec. \ref{sub:Mappingc}) onto fermionic correlators, so that
we can derive in Sect. \ref{sec:adl} the precise form of their asymptotic
behaviour at large distances by exploiting the machinery of Toeplitz
determinants. Sect. \ref{sec:DMRG} reports a comparison with DMRG
simulations of the system in equation (\ref{eq:HlamD}) while Sect.
\ref{sec:Conc} is devoted to conclusions.

\section{Mapping onto spinless fermions\label{sec:Mapping}}

The basic idea underlying the approximation used in this work is the
spin solid picture of the Néel state(s):\begin{equation}
|N\rangle=|\uparrow\downarrow\uparrow\cdots\downarrow\uparrow\downarrow\cdots\uparrow\downarrow\rangle\label{eq:Neelstate}\end{equation}
 which is, in fact, the GS of the Hamiltonian (\ref{eq:HlamD}) for
$\lambda\to\infty$ at fixed $D$ or $D\to-\infty$ and $\lambda>0$.
Actually the GS is doubly degenerate: for a given configuration of
the type (\ref{eq:Neelstate}) with, say, $S_{i}^{z}=1$ at the reference
site $i=0$, the energy is unchanged by the Z$_{2}$ transformation
$T=\exp({\rm i}\pi\sum_{j}S_{j}^{y})$ that performs a $\pi$-rotation
about the $y$-axis (spin-flip). We will refer to $\vert N\rangle$
and $\vert\bar{N}\rangle=T\vert N\rangle$ as Néel and anti-Néel states,
respectively. Now, in a perturbative fashion, when $\vert D\vert$
and/or $\lambda\gg1$ the effect of the transverse terms in the Hamiltonian
$S_{i}^{x,y}S_{i+1}^{x,y}$ is to:

\begin{description}
\item [{i)}] create pairs of adjacent sites with $S_{i}^{z}=S_{i+1}^{z}=0:$
$|\uparrow\downarrow\rangle\to\vert0\;0\rangle$;
\item [{ii)}] move the zeroes in the AFM background, e.g.: $|\uparrow0\rangle\to\vert0\uparrow\rangle$; 
\item [{iii)}] re-create a pair $\uparrow\downarrow$ or $\downarrow\uparrow$
in place of a pair of adjacent zeroes.
\end{description}
Notice that ii) preserves the AFM order, albeit not on nearest neighbours
but mediated by string of zeroes (hidden order). Again, due to the
AFM order (induced by $\lambda>0$ and by the transverse terms), even
if both states of iii) can be created in an {}``island'' of zeroes,
as far as the low-energy part of the spectrum is concerned, one of
the two will be preferred according to the orientation of the surrounding
spins, that is, by the hidden AFM order. Note also that $\vert N\rangle$
and $\vert\bar{N}\rangle$ are connected through a large number of
virtual processes, so that in the thermodynamic limit only one of
the two will be selected by a spontaneous symmetry breaking mechanism
induced by an infinitesimal staggered magnetic field. Alternatively,
if the system under consideration is described by a thermal density
matrix $\exp(-\beta H)$, when $\beta\to\infty$ the GS reduces to
a symmetric mixed state $\vert N\rangle\langle N\vert+\vert\bar{N}\rangle\langle\bar{N}\vert$.

Once the question of the GS is accounted for, from the scenario above
one can see that the orientation of spins with nonzero component along
$z$ is determined by the hidden order and can be taken for granted.
The validity of such an approximation is ultimately measured by the
values of the longitudinal SOP: the closer is ${\cal O}_{z}$ to unity
the higher is the AFM order of nonzero spins. We then introduce the
following fermionic picture: assign a spinless fermion $\vert+_{i}\rangle\equiv c_{i}^{\dagger}\vert-_{i}\rangle$
at site $i$ if $S_{i}^{z}\neq0$ and no fermions $\vert-_{i}\rangle\equiv\vert0_{i}\rangle$
in the spin language if $S_{i}^{z}=0$. (This notation for spinless
fermions has a direct translation in the language of the XY model
that will be introduced at the end of this section.) Process ii) is
nothing but a hopping of spinless fermions, while processes i) and
iii) represent annihilation and creation of pairs $\vert+_{i}+_{i+1}\rangle$.
The density of nonzero spins $(S_{i}^{z})^{2}$ is simply translated
to the local fermion number $n_{i}=c_{i}^{\dagger}c_{i}$, while,
due to the underlying AFM order, the Ising-like term takes the form
$-\lambda n_{i}n_{i+1}$ that contributes with a negative energy when
two fermions are present on adjacent sites.

Hence, under the hypothesis of hidden AFM order, the dynamics of equation
(\ref{eq:HlamD}) is reproduced by the following effective fermionic
model\begin{equation}
H_{{\rm f}}=\sum_{j}\left(c_{j}^{\dagger}c_{j+1}+c_{j+1}^{\dagger}c_{j}+c_{j}^{\dagger}c_{j+1}^{\dagger}+c_{j+1}c_{j}-\lambda n_{j}n_{j+1}+Dn_{j}\right)\label{eq:Hfermionic}\end{equation}
 (acting in a reduced Hilbert space ${\cal H}=\otimes_{i}{\cal H}_{i}^{(2)}$where
${\cal H}^{(2)}$ denotes the local Hilbert space of a two-level system
- as that of a spinless fermion or a spin-1/2 introduced below). It
should be observed that equation (\ref{eq:Hfermionic}) with $D=0$
is essentially equivalent (apart from an additive constant) to equation
(2) of ref. \cite{g1989} once a particle-hole transformation $n_{i}\to1-n_{i}$
is performed at every site.

Following G\'{o}mez-Santos \cite{g1989} we now proceed to a further
approximation on the fermionic Hamiltonian that is not amenable to
an exact treatment due to the $\lambda$-term. At the Hartree-Fock
level this term can be approximated as:\[
n_{j}n_{j+1}\simeq\left(n_{j}+n_{j+1}\right)\langle n_{j}\rangle-\left(c_{j}^{\dagger}c_{j+1}\langle c_{j+1}^{\dagger}c_{j}\rangle+{\rm h.c.}\right)+\left(c_{j}^{\dagger}c_{j+1}^{\dagger}\langle c_{j+1}c_{j}\rangle+{\rm h.c.}\right)\]
\[
-\left(\langle n_{j}\rangle^{2}-\langle c_{j+1}^{\dagger}c_{j}\rangle\langle c_{j}^{\dagger}c_{j+1}\rangle+\langle c_{j+1}c_{j}\rangle\langle c_{j}^{\dagger}c_{j+1}^{\dagger}\rangle\right)\]
 where the expectation values $\langle\dots\rangle$ now are taken
with respect to the GS of the quadratic Hamiltonian\[
H_{{\rm HF}}=\sum_{j}\left[\left(1+\lambda A\right)c_{j}^{\dagger}c_{j+1}+\left(1-\lambda B\right)c_{j}^{\dagger}c_{j+1}^{\dagger}+{\rm h.c.}\right]\]
\begin{equation}
+\left(D-2\lambda n_{0}\right)n_{j}+\lambda\left(n_{0}^{2}-|A|^{2}+|B|^{2}\right)\label{eq:HHF}\end{equation}
 where the parameters\[
n_{0}\equiv\langle n_{j}\rangle\;,\;\; A\equiv\langle c_{j+1}^{\dagger}c_{j}\rangle\;,\;\; B=\langle c_{j+1}c_{j}\rangle\]
 have to be determined self-consistently. The advantage of a Hamiltonian
of the form (\ref{eq:HHF}) is that it can be diagonalized by means
of a Bogoliubov transformation\[
\eta_{k}=\cos\frac{\theta_{k}}{2}c_{k}+{\rm i}\sin\frac{\theta_{k}}{2}c_{-k}^{\dagger}\]
 where $c_{k}=1/\sqrt{L}\sum_{j}c_{j}\exp(-{\rm i}jk)$ and $\theta_{k}$
is given by\[
e^{{\rm i}\theta_{k}}=\frac{\left(\cos k-h+{\rm i}\gamma\sin k\right)}{\Lambda_{k}}\]
 where\begin{equation}
h\equiv\frac{2\lambda n_{0}-D}{2\left(1+\lambda A\right)}\;,\;\;\gamma\equiv\frac{1-\lambda B}{1+\lambda A}\label{eq:hgamma}\end{equation}
 \[
\Lambda_{k}=\sqrt{\left(\cos k-h\right)^{2}+\gamma^{2}\sin k^{2}}\]
Note that, as we are interested in the thermodynamic limit, we do
not specify here the boundary conditions on the spin and fermionic
Hamiltonians. The momenta are quantized as $\Delta k=2\pi/L$ and
their precise location within the first Brillouin zone depend on the
conditions imposed on the end sites. However, for $L\to\infty$\[
\frac{1}{L}\sum_{k}\to\frac{1}{2\pi}\int_{0}^{2\pi}{\rm d}k.\]
 Apart from additive terms of $O(L^{-1})$ the Hamiltonian in diagonal
form is\[
H_{{\rm HF}}=2(1+\lambda A)\sum_{k}\Lambda_{k}\left(\eta_{k}^{\dagger}\eta_{k}-\frac{1}{2}\right)+U\]
 and $U=(D-2\lambda n_{0})/2+\lambda(n_{0}^{2}-A^{2}+B^{2})$. In
the thermodynamic limit, the self-consistent equations are\begin{equation}
n_{0}=\frac{1}{2}-\frac{1}{2\pi}\int_{0}^{\pi}dk\frac{-h\left(n_{0},A\right)+\cos k}{\Lambda(k)}\label{eq:ien0}\end{equation}
 \begin{equation}
A=-\frac{1}{2\pi}\int_{0}^{\pi}dk\frac{\left(-h\left(n_{0},A\right)+\cos k\right)\cos k}{\Lambda(k)}\label{eq:ieA}\end{equation}
 \begin{equation}
B=-\frac{1}{2\pi}\int_{0}^{\pi}dk\frac{\gamma\left(A,B\right)\sin^{2}k}{\Lambda(k)}.\label{eq:ieB}\end{equation}

The notation used in equation (\ref{eq:hgamma}) is the one commonly
used for the XY spin-1/2 model in a transverse field. In fact, by
(inverse) Jordan-Wigner transform one gets \cite{h1999}\begin{equation}
H_{{\rm HF}}\rightarrow H_{{\rm XY}}=\sum_{j}\left(\frac{1+\gamma}{2}\right)\sigma_{j}^{x}\sigma_{j+1}^{x}+\left(\frac{1-\gamma}{2}\right)\sigma_{j}^{y}\sigma_{j+1}^{y}-h\sigma_{j}^{z}\label{eq:HXY}\end{equation}
 where the $\sigma_{j}^{\alpha}$'s are Pauli matrices at site $j$.
This model is known to be critical at $h=\pm1$ for $\gamma\neq0$,
where it belongs to the $c=1/2$ universality class (the same as the
2D classical Ising model) and at $\gamma=0$ for $h\in(-1,1)$ where
the universality class becomes that of the compactified free boson,
$c=1$.

From the numerical solutions of (\ref{eq:ien0})-(\ref{eq:ieB}) it
turns out that in the Haldane and Néel phases $A<0$ and $B<0$ so
that $\gamma>1$ (as long as $\lambda\vert A\vert<1$ - some representative
cases are listed in table \ref{tab:n0ABhgamma}), while most studies
are limited to $\vert\gamma\vert<1$. As a consequence the region
$\gamma^{2}<1-h^{2}$ corresponding to oscillations with wavenumber
different from $\pi$ \cite{bm1971,h1999} is not present in our case.
However, having $\gamma>1$ does not affect the critical condition
we are interested in, that remains $h=\pm1$. In these cases we have
just $c=1/2$, as reported before \cite{chs2003} for the Haldane-to-Néel
transition. At $\lambda=0$, $D\cong-2$ this transition line merges
with the boundary towards the so-called XY phases corresponding to
$c=1$. Interestingly this change of universality class is captured
also by our approximation since for $\lambda=0$, $D=-2$ the self-consistent
solution yields just $\gamma=1$ and $h=1$. From the data in table
\ref{tab:n0ABhgamma} one can also estimate, for example, the critical
value of $D$ at fixed $\lambda=1$; the result is $D_{{\rm c}}\cong-0.214$,
which is not in good quantitative agreement with the numerical value
$D_{{\rm c}}=-0.315$ \cite{chs2003,debeor2003}. The perturbation
of the isotropic Heisenberg Hamiltonian with $\lambda>1$ and $D=0$,
instead, seems to be better described by the spinless fermions approach;
already at this level of approximation the value $\lambda_{{\rm c}}=1.125$
found in \cite{g1989} is close to our best DMRG independent estimate
$\lambda_{{\rm c}}=1.1856$ \cite{debo2004}. Even if it is likely
that the inclusion of configurations with nearest-neighbour parallel
spins could improve the results, as discussed by G\'{o}mez-Santos
\cite{g1989}, we do not insist along this line here because we are
ultimately interested in the decay laws of correlation functions that
are essentially dictated by the universality classes. In fact, it
is important to stress that neither the extension to $D\neq0$, nor
the extension of the model as in equation (8) of ref. \cite{g1989}
modify the universality class of the transition, that remains of the
$c=1/2$ (or Ising) type for $\lambda>0$. Although the location of
the critical points and of the prefactors depend on the values of
the parameters, the scaling dimensions of the operators in the continuum
field theory (i.e. the decay exponents of the correlation functions)
do not change when we move along the $c=1/2$ line. Nonetheless, due
to the lack of an explicit mapping of the spin-1 strings onto the
corresponding correlators in the Ising fermionic field theory, up
to now the exponents appearing in the large-distance decay of string
correlation functions were unknown. This is precisely the subject
of subsecs. \ref{sub:SOPz} and \ref{sub:SOPx}. Eventually, we note
that alternative pictures of the Haldane gap in fermionic language
can be derived by perturbation theory near the Babujian-Takhtajan
integrable biquadratic spin-1 chain \cite{t1990} or from two-leg
ladders with ferromagnetic coupling on the rungs \cite{snt1996}.

\begin{table}

\caption{Self-consistent estimates of the three decoupling parameters $n_{0}$,
$A$ and \textbf{$B$} of equations (\ref{eq:ien0})-(\ref{eq:ieB})
for some choices of $\lambda$ and $D$ in the Haldane and Néel phases.
It must be kept in mind that the continuum versions of the self-consistent
equations neglect some $O(L^{-1})$ terms coming from isolated contributions
at wavenumber 0 or $\pi$. Last two column contain the corresponding
parameters $h$ and $\gamma$ of the effective XY model according
to equation (\ref{eq:hgamma}).\label{tab:n0ABhgamma}}

\begin{tabular}{@{}lllllll}
\hline 
$\lambda$&
$D$&
$n_{0}$&
$A$&
$B$&
$h$&
$\gamma$\tabularnewline
\hline 
$1$&
$0$&
$0.709$&
$-0.158$&
$-0.253$&
$0.842$&
$1.49$\tabularnewline
$1$&
$-0.125$&
$0.745$&
$-0.137$&
$-0.246$&
$0.936$&
$1.44$\tabularnewline
$1$&
$-0.200$&
$0.774$&
$-0.117$&
$-0.240$&
$0.990$&
$1.41$\tabularnewline
$1$&
$-0.250$&
$0.800$&
$-0.0979$&
$-0.235$&
$1.03$&
$1.37$\tabularnewline
$1$&
$-0.300$&
$0.816$&
$-0.0866$&
$-0.231$&
$1.06$&
$1.35$\tabularnewline
$1$&
$-0.315$&
$0.820$&
$-0.0837$&
$-0.230$&
$1.07$&
$1.34$\tabularnewline
$1$&
$-0.330$&
$0.824$&
$-0.0811$&
$-0.229$&
$1.08$&
$1.34$\tabularnewline
$1$&
$-0.345$&
$0.828$&
$-0.0786$&
$-0.228$&
$1.09$&
$1.33$\tabularnewline
$1$&
$-0.400$&
$0.841$&
$-0.0706$&
$-0.223$&
$1.12$&
$1.32$\tabularnewline
$1$&
$-0.450$&
$0.850$&
$-0.0645$&
$-0.219$&
$1.15$&
$1.30$\tabularnewline
$1$&
$-0.750$&
$0.893$&
$-0.0406$&
$-0.198$&
$1.32$&
$1.25$\tabularnewline
$1$&
$-0.875$&
$0.904$&
$-0.0344$&
$-0.190$&
$1.39$&
$1.23$\tabularnewline
$1$&
$-10$&
$0.996$&
$-0.000317$&
$-0.0433$&
$6.00$&
$1.04$\tabularnewline
$5$&
$-0.125$&
$0.991$&
$-0.000853$&
$-0.0649$&
$5.04$&
$1.33$\tabularnewline
\hline
\end{tabular}

\end{table}

At this stage it is interesting to compare the entanglement properties
of the original spin-1 model (eq. (\ref{eq:HlamD})) with those of
the XY spin-1/2 chain resulting from the mapping. On the one hand,
for the former it has been shown \cite{cvdebm2006} that at the isotropic
Heisenberg point $\lambda=1$, $D=0$ there is long-distance spin-1
(qutrit) entanglement in the thermodynamic limit for two sites arbitrarily
far apart. It is reasonable to expect that this entanglement survives
in a neighbourhood of the isotropic point. On the other hand, in ref.
\cite{oetal2002} it is stated that the qubit entanglement in the
XY model with transverse field vanishes beyond a distance of order
$\gamma^{-1}$. In our case $\gamma>1$ and the degrees of freedom
of the qubits represent the presence or the absence of an effective
particle with ${\vert S}_{i}^{z}\vert=1$. Therefore we are led to
speculate that wherever there is full spin-1 entanglement in the vicinity
of the Heisenberg point, this is due to the spin correlations between
the sites with $S_{i}^{z}\neq0$. Recalling the hypothesis of underlying
string order and imagining to eliminate the sites with $S_{i}^{z}=0$,
the qualitative picture of the long-distance entangled states in the
Haldane region if that of a Greenberger-Horne-Zeilinger state \cite{ghz1989}
with effective AFM order $\vert\dots\uparrow\downarrow\uparrow\downarrow\dots\rangle+\vert\dots\downarrow\uparrow\downarrow\uparrow\dots\rangle$
.

\subsection{Mapping for the spin-spin and string correlators\label{sub:Mappingc}}

We shall exploit now the mapping from spin-1 to spinless fermions,
based on the existence of an underlying string order, to translate
the various spin-1 correlation functions onto expectation values of
strings of fermionic operators that can be computed exactly when the
Hamiltonian has the form (\ref{eq:HHF}). Let us start from the $z$-component
of the spin\begin{equation}
S_{j}^{z}\to n_{j}K_{j}\to\frac{1+\sigma_{j}^{z}}{2}K_{j}\label{eq:Sz}\end{equation}
 where $K_{j}=\exp({\rm i}\pi\sum_{i<j}n_{i})=\prod_{i<j}(-\sigma_{i}^{z})$
is a Jordan-Wigner tail that accounts for the correct sign when $S_{j}^{z}\neq0$
assuming, conventionally, that the first nonzero spin is pointing
up. By inserting the expression $S_{j}^{z}=\frac{1+\sigma_{j}^{z}}{2}\prod_{i<j}(-\sigma_{j}^{z})$
into the definition of the longitudinal spin-spin correlation function
and using the properties of Pauli matrices one finds

\begin{eqnarray}
\mathcal{C}_{z}(R) & \to & \frac{1}{4}(-1)^{R}\langle\left(1+\sigma_{j}^{z}\right)\prod_{k<j}\left(-\sigma_{k}^{z}\right)\prod_{k<j+R}\left(-\sigma_{k}^{z}\right)\left(1+\sigma_{j+R}^{z}\right)\rangle\nonumber \\
 & = & \frac{1}{4}\left(\langle\prod_{k=j}^{j+R}\sigma_{k}^{z}\rangle+\langle\prod_{k=j+1}^{j+R}\sigma_{k}^{z}\rangle+\langle\prod_{k=j}^{j+R-1}\sigma_{k}^{z}\rangle+\langle\prod_{k=j+1}^{j+R-1}\sigma_{k}^{z}\rangle\right).\label{eq:Cz-sigma}\end{eqnarray}

As far as the transverse correlation functions are concerned, it can
be checked by direct inspection on a generic configuration with perfect
string order that the identification\begin{equation}
{\cal C}_{x}(R)=\frac{1}{2}\left(S_{j}^{+}S_{j+R}^{-}+S_{j}^{-}S_{j+R}^{+}\right)\to\sigma_{j}^{x}\prod_{k=j+1}^{j+R-1}\left(\frac{1-\sigma_{k}^{z}}{2}\right)\sigma_{j+R}^{x}\label{eq:S+S--sigma}\end{equation}
 has the correct action, since the only cases in which the l.h.s.
does not break the string order are those with $S_{k}^{z}=0$, that
is $\sigma_{k}^{z}=-1$, on all sites between $j$ and $j+R$. The
product on the r.h.s. of (\ref{eq:S+S--sigma}) is exactly the expression
involved in the so-called emptiness formation probability (see, for
example, \cite{fa2005} and refs. therein).

Let us now study the spin-1 strings. Along the $z$-direction we have
simply\begin{equation}
e^{i\pi\sum_{k<j}S_{k}^{z}}=\prod_{k<j}\left(1-2(S_{k}^{z})^{2}\right)\to\prod_{k<j}\left(-\sigma_{k}^{z}\right).\label{eq:expSz-sigma}\end{equation}
 Again by using the relation $S_{j}^{z}=\frac{1+\sigma_{j}^{z}}{2}\prod_{i<j}(-\sigma_{j}^{z})$
and plugging the string written above into eq. (\ref{eq:scf}) one
gets \begin{eqnarray}
{\cal O}_{z}(R)\to\left\langle \left(\frac{1+\sigma_{j}^{z}}{2}\right)\prod_{k<j}\left(-\sigma_{k}^{z}\right)\prod_{k=j+1}^{j+R-1}\left(-\sigma_{k}^{z}\right)\prod_{k<j+R}\left(-\sigma_{k}^{z}\right)\left(\frac{1+\sigma_{j+R}^{z}}{2}\right)\right\rangle \\
=-\langle\left(\frac{1+\sigma_{j}^{z}}{2}\right)\left(\frac{1+\sigma_{j+R}^{z}}{2}\right)\rangle=-\frac{1}{4}\left(1+\langle\sigma_{j}^{z}\rangle+\langle\sigma_{j+R}^{z}\rangle+\langle\sigma_{j}^{z}\sigma_{j+R}^{z}\rangle\right)\label{eq:Oz-sigma}\end{eqnarray}
Note that in the language of the effective XY model, the Néel correlation
function (\ref{eq:Cz-sigma}) involves a string of Pauli operators
whereas the string correlation function (\ref{eq:Oz-sigma}) involves
only one- and two-points correlators of the $\sigma$'s. Thanks to
equation (\ref{eq:expSz-sigma}) we easily obtain also the pure-string
correlation function as:\begin{equation}
G_{H}(R)\to(-1)^{R+1}\left\langle \prod_{j=i}^{i+R}\sigma_{j}^{z}\right\rangle .\label{eq:GH-sigma}\end{equation}
 From equation (\ref{eq:Cz-sigma}) we see that, in this approach,
$G_{H}(R)$ is nothing but the first term of the usual spin-spin correlation
function ${\cal C}_{z}(R)$ apart from the prefactor.

\textcolor{black}{Along the $x$-direction, instead, we exploit the
fact that\[
{\rm e}^{{\rm i}\pi S^{x}}=\left(\begin{array}{ccc}
0 & 0 & -1\\
0 & -1 & 0\\
-1 & 0 & 0\end{array}\right)\]
 that is, apart from an overall sign, the operator above performs
a swap between $S_{j}^{z}=1$ and $S_{j}^{z}=-1$ leaving the $S_{j}^{z}=0$
component isolated. This swap is important because it can be checked
by direct inspection that, for every possible combination of the spin
at sites $j$ and $j+R$ that respects the hidden AFM order, both
with an even or an odd number of nonzero spins in between, the action
of ${\cal O}_{x}(R)$ produces only one allowed configuration and
some other forbidden ones. More precisely, using the spin-1/2 operator
$\sigma^{x}$ that changes empty sites into occupied sites and viceversa
we can write\[
\mathcal{O}_{x}(R)\to(-1)^{R-1}\frac{\langle\sigma_{j}^{x}\sigma_{j+R}^{x}\rangle}{2}\]
 where the inner spin-1 transverse string contributes with the sign
prefactors. Thanks to hidden order in our reduced Hilbert space, the
spin-1/2 configurations generated by $\sigma^{x}$ represent the allowed
spin-1 states and the forbidden ones are automatically filtered out.
The coefficient $1/2$ comes from the matrix elements of $S^{x}$
at sites $j$ and $j+R$.}

\textcolor{black}{Now, thanks to the fact that the Hamiltonian (\ref{eq:HHF})
is quadratic in the fermionic operators, all the correlation functions
can be evaluated using Wick's theorem. Following the notation of the
seminal paper by Lieb, Schultz and Mattis \cite{lsm1961} we introduce
the operators $A_{j}=c_{j}^{\dagger}+c_{j}$ and $B_{j}=c_{j}^{\dagger}-c_{j}$
that allow to express the basic two-point correlations as\[
\langle\sigma_{l}^{x}\sigma_{m}^{x}\rangle=\langle B_{l}A_{l+1}B_{l+1}\cdots A_{m-1}B_{m-1}A_{m}\rangle\]
 \[
\langle\sigma_{l}^{z}\sigma_{m}^{z}\rangle=\langle A_{l}B_{l}A_{m}B_{m}\rangle\]
 with $Q_{lm}\equiv\langle A_{l}A_{m}\rangle=\delta_{lm}$ and $S_{lm}=\langle B_{l}B_{m}\rangle=-\delta_{lm}$.
If we further assume translational invariance (i.e. PBC) along the
chain we have\begin{equation}
\langle\sigma_{l}^{x}\sigma_{m}^{x}\rangle=\left|\begin{array}{cccc}
G_{-1} & G_{-2} & \cdots & G_{l-m}\\
\vdots &  &  & \vdots\\
G_{m-l-2} & \cdots &  & G_{-1}\end{array}\right|\label{eq:sigmaxx}\end{equation}
 \begin{equation}
\langle\sigma_{l}^{z}\sigma_{m}^{z}\rangle=G_{0}^{2}-G_{m-l}G_{l-m}\label{eq:sigmazz}\end{equation}
 where $G_{-R}\equiv\langle B_{j}A_{j+R}\rangle=-\langle A_{j+R}B_{j}\rangle$.
In particular, $G_{0}=\langle(c_{j}^{\dagger}-c_{j})(c_{j}^{\dagger}+c_{j})\rangle=2\langle n_{j}\rangle-1=\langle\sigma_{j}^{z}\rangle$,
independent of $j$ and $\langle\sigma_{j}^{z}\sigma_{j+R}^{z}\rangle=\langle\sigma_{j}^{z}\rangle^{2}-G_{R}G_{-R}$.
The $R$-dependence of ${\cal O}_{z}(R)$ and ${\cal O}_{x}(R)$ is
given directly by $\langle\sigma_{j}^{z}\sigma_{j+R}^{z}\rangle$
and $\langle\sigma_{j}^{x}\sigma_{j+R}^{x}\rangle$ respectively.
The ordinary correlators ${\cal C}_{x,z}(R)$ require a step more
since they involve strings of Pauli operators. For example, each of
the terms in equations (\ref{eq:Cz-sigma}) and (\ref{eq:GH-sigma})
has the form $\langle\prod_{k}B_{k}A_{k}\rangle$. When $R\to\infty$
all the four terms in equation (\ref{eq:Cz-sigma}) tend to coincide
so that\[
{\cal C}_{z}(R)\simeq(-1)^{R+1}G_{H}(R)=\langle B_{j}A_{j}B_{j+1}A_{j+1}\cdots B_{j+R-1}A_{j+R-1}B_{j+R}A_{j+R}\rangle.\]
 Exploiting Wick's theorem, Caianiello and Fubini \cite{cf1952} have
shown that the expectation value above can be expressed as a Pfaffian\[
\left.\begin{array}{cccccccccc}
{\rm Pf}|S_{-1} & S_{-2} & \cdots & S_{-R+1} & S_{-R} & G_{0} & G_{-1} & \cdots & G_{-R+1} & G_{-R}\\
 & S_{-1} & \cdots & S_{-R+2} & S_{-R+1} & G_{1} & G_{0} & \cdots & G_{-R+2} & G_{-R+1}\\
 &  & \ddots & \vdots & \vdots & \vdots & \vdots & \vdots & \vdots & \vdots\\
 &  &  & S_{-1} & S_{-2} & G_{R-2} & G_{R-3} & \cdots & G_{-1} & G_{-2}\\
 &  &  &  & S_{-1} & G_{R-1} & G_{R-2} & \cdots & G_{0} & G_{-1}\\
 &  &  &  &  & G_{R} & G_{R-1} & \cdots & G_{1} & G_{0}\\
 &  &  &  &  &  & Q_{-1} & \cdots & Q_{-R+1} & Q_{-R}\\
 &  &  &  &  &  &  & \ddots & \vdots & \vdots\\
 &  &  &  &  &  &  &  & Q_{-1} & Q_{-2}\\
 &  &  &  &  &  &  &  &  & Q_{-1}\end{array}\right|\]
 Thanks to the fact that $Q_{l\neq m}=S_{l\neq m}=0$ this Pfaffian
reduces to a Toeplitz determinant \cite{vd1999} and we get\begin{equation}
{\cal C}_{z}(R)=\left|\begin{array}{ccccc}
{-G}_{0} & {-G}_{-1} & \cdots & {-G}_{-R+1} & {-G}_{-R}\\
{-G}_{1} & -G_{0} & \cdots & {-G}_{-R+2} & {-G}_{-R+1}\\
\vdots & \vdots & \cdots & \vdots & \vdots\\
{-G}_{R-1} & {-G}_{R-2} & \cdots & {-G}_{0} & {-G}_{-1}\\
{-G}_{R} & -G_{R-1} & \cdots & {-G}_{1} & -G_{0}\end{array}\right|\label{eq:Cz-Toeplitz}\end{equation}
 So, the determinants of the matrices with entries $G_{j}$ becomes
the central quantities of our analysis.}

The matter is more complicated for the transverse spin-spin correlator
essentially because it eventually involves a Toeplitz determinant
generated by a matrix-valued symbol that may also become singular.
According to ref. \cite{l2007} this case is not yet solved in the
theory of Toeplitz determinants and in ref. \cite{ijk2007} it has
been suggested to extend directly the procedure valid in the nonsingular
case. Fortunately in our case a workaround is possible: thanks to
a suitable diagonalization, we are able to complete the calculation
of the dominant contribution to ${\cal C}_{x}(R)$ in terms of a product
of Toeplitz determinants, each one computed using the Fisher-Hartwig
conjecture (see, for instance, App. A.2 in \cite{fa2005}). The details
of this procedure are reported in the Appendix.

\section{Asymptotic decay laws\label{sec:adl}}

\subsection{Longitudinal string correlation ${\cal O}_{z}(R)$\label{sub:SOPz}}

The first object we will compute is the longitudinal string correlator.
From equations (\ref{eq:Oz-sigma}) and (\ref{eq:sigmazz}) we get\[
{\cal O}_{z}(R)=-\frac{1}{4}\left[\left(1+\langle\sigma_{j}^{z}\rangle\right)^{2}-G_{R}G_{-R}\right].\]
 Following Barouch and McCoy \cite{bm1971} (in particular their eq.
(6.12)) we express $G_{R}$ as follows\begin{equation}
G_{R}=-\frac{1}{2\pi}\int_{0}^{2\pi}dk{\rm e}^{-{\rm i}k(R+1)}\left[\frac{\left(1-\lambda_{1}^{-1}{\rm e}^{{\rm i}k}\right)\left(1-\lambda_{2}^{-1}{\rm {\rm e}}^{{\rm i}k}\right)}{\left(1-\lambda_{1}^{-1}{\rm e}^{-{\rm i}k}\right)\left(1-\lambda_{2}^{-1}{\rm e}^{-{\rm i}k}\right)}\right]^{1/2}=-\frac{1}{2\pi}\int_{0}^{2\pi}dk{\rm e}^{-{\rm i}kR}c({\rm e}^{{\rm i}k})\label{eq:GR}\end{equation}
 with\begin{equation}
\lambda_{1,2}=\frac{h\pm\sqrt{h^{2}-(1-\gamma^{2})}}{1-\gamma}.\label{eq:lambda12}\end{equation}
 and\[
c({\rm e}^{{\rm i}k})={\rm e}^{-{\rm i}k}\sqrt{\frac{(1-\lambda_{1}^{-1}{\rm e}^{{\rm i}k})(1-\lambda_{2}^{-1}{\rm e}^{{\rm i}k})}{(1-\lambda_{1}^{-1}{\rm e}^{-{\rm i}k})(1-\lambda_{2}^{-1}{\rm e}^{-{\rm i}k})}}.\]
 Note that since $\gamma>1$ the two roots of the numerator are always
real; the behaviour for $R\to\infty$ is controlled by $\lambda_{2}$.
From equations (6.17), (6.14) and (6.18) in ref. \cite{bm1971} we
have, respectively:

\begin{itemize}
\item \textit{Haldane phase $h<1$}($\lambda_{2}>1$)\[
{\cal O}_{z}(R)\simeq{\cal O}_{z}+\frac{1}{8\pi}\frac{{\rm e}^{-2R/\xi}}{R^{2}}\;,\;\;\xi\equiv1/\ln\lambda_{2}\]
 
\item \textit{Critical line} $h=1$ ($\lambda_{2}=1$)\[
{\cal O}_{z}(R)\simeq{\cal O}_{z}+\frac{1}{4\pi^{2}}\frac{1}{R^{2}}\]

\item \textit{Néel phase $h>1$} ($0<\lambda_{2}<1$)\[
{\cal O}_{z}(R)\simeq{\cal O}_{z}+\frac{1}{8\pi}\frac{{\rm e}^{-2R/\xi}}{R^{2}}\;,\;\;\xi\equiv-1/\ln\lambda_{2}\]

\end{itemize}
In every case the asymptotic value ${\cal O}_{z}\neq0$ is simply
interpreted as a non-saturated value of the magnetization along $z$
in the XY model in transverse field \begin{equation}
{\cal O}_{z}=-\frac{\left(1+\langle\sigma_{j}^{z}\rangle\right)^{2}}{4},\label{eq:Oz-vs-sigmaz}\end{equation}
 where $\langle\sigma_{j}^{z}\rangle=G_{0}(h,\gamma)$ can be computed
using equation (\ref{eq:GR}) at $R=0$.

\subsection{Longitudinal spin-spin correlation function ${\cal C}_{z}(R)$ and
pure string correlator $G_{H}(R)$\label{sub:CzGH}}

The asymptotic behaviour of the Toeplitz determinant in equation (\ref{eq:Cz-Toeplitz})
can be found using the same technique as in \cite{w1966}, since (apart
from a sign) the generating symbol $c({\rm e}^{{\rm i}k})$ is essentially
the same used by Wu. Then we find:

\begin{itemize}
\item \textit{Haldane phase $h<1$} ($\lambda_{2}>1$)\[
{\cal C}_{z}(R)\simeq(-1)^{R+1}G_{H}(R)=\frac{1}{\sqrt{\pi}}(1-\lambda_{1}^{-2})^{1/4}(1-\lambda_{2}^{-2})^{-1/4}(1-\lambda_{1}^{-1}\lambda_{2})^{-1/2}\frac{{\rm e}^{-R/\xi}}{R^{1/2}}\]
which corresponds to the known decay behaviour at the isotropic Heisenberg
point, as predicted by the nonlinear $\sigma$-model approach (see,
for example, \cite{cvetal2002}). Moreover, in refs. \cite{ldqsy1999,empr2000}
it was argued that the same behaviour of the connected longitudinal
correlation function persists also in presence of a staggered magnetic
field; in this sense such a behaviour could be considered a signature
of the Haldane phase, robust against anisotropic perturbations.
\item \textit{Critical line} $h=1$ ($\lambda_{2}=1$)\[
{\cal C}_{z}(R)\simeq(-1)^{R+1}G_{H}(R)={\rm e}^{1/4}2^{1/12}{\cal A}^{-3}\frac{1}{{(\gamma R)}^{1/4}}\]
 where ${\cal A}=1.282427130\dots$ denotes Glaisher's constant \cite{w1966}.
\item \textit{Néel phase $h>1$} ($0<\lambda_{2}<1$)\[
{\cal C}_{z}(R)\simeq(-1)^{R+1}G_{H}(R)=(1-\lambda_{1}^{-2})^{1/4}(1-\lambda_{2}^{2})^{1/4}(1-\lambda_{1}^{-1}\lambda_{2})^{-1/2}\left[1+\frac{1}{2\pi(\lambda_{2}^{-1}-\lambda_{2})^{2}}\frac{{\rm e}^{-2R/\xi}}{R^{2}}\right]\]

\end{itemize}
Apart from the nonzero asymptotic value for $h>1$, that serves as
an order parameter for the Néel phase (the ordered phase $T<T_{{\rm c}}$
in Wu's paper \cite{w1966}), it must be noticed that both the power
of $R$ in the denominator and the exponential constant are different
on the two sides of the transition. The roots $\lambda_{1,2}$ and
the bulk correlation length $\xi$ are the same as in subsec. \ref{sub:SOPz}
(see eq. (\ref{eq:lambda12})).

\subsection{Transverse string correlation function ${\cal O}_{x}(R)$\label{sub:SOPx}}

\begin{itemize}
\item \textit{Haldane phase $h<1$} ($\lambda_{2}>1$). The nonzero asymptotic
value ${\cal O}_{x}$ comes from the long-range order $\lim_{R\to\infty}\langle\sigma_{0}^{x}\sigma_{R}^{x}\rangle$
in the XY model with spontaneous breaking of the symmetry $\sigma^{x}\to-\sigma^{x}$.
The result can be borrowed directly from equation (4.1) of \cite{bm1971}\begin{equation}
{\cal O}_{x}(R)\simeq-\frac{\left[\gamma^{2}(1-h^{2})\right]^{1/4}}{1+\gamma}\left[1+\frac{1}{2\pi R^{2}}\frac{{\rm e}^{-2R/\xi}}{(\lambda_{2}-\lambda_{2}^{-1})^{2}}\right]\label{eq:Ox-Haldane}\end{equation}
 with $\xi$ having the same meaning of subsec. \ref{sub:SOPz}.
\item \textit{Critical line} $h=1$ ($\lambda_{2}=1$). There is no long-range-order
in $\langle\sigma_{0}^{x}\sigma_{R}^{x}\rangle$, that decays to zero
as $R^{-1/4}$ as expected from the scaling dimension $1/8$ of the
primary operator in the $c=1/2$ CFT \cite{h1999}. Using equation
(4.7) in \cite{bm1971} we have \begin{equation}
{\cal O}_{x}(R)\simeq-\frac{\gamma}{1+\gamma}{\rm e}^{1/4}2^{1/12}{\cal A}^{-3}\frac{1}{(\gamma R)^{1/4}}\label{eq:Ox-crit}\end{equation}

\item \textit{Néel phase $h>1$} ($0<\lambda_{2}<1$). Equation (4.25) in
ref. \cite{bm1971} is translated to\begin{equation}
{\cal O}_{x}(R)\simeq-\frac{1}{2\sqrt{\pi}}\frac{{\rm e}^{-R/\xi}}{R^{1/2}}\left[(1-\lambda_{2}^{2})^{-1}(1-\lambda_{1}^{-2})(1-\lambda_{1}^{-1}\lambda_{2}^{-1})^{2}\right]^{1/4}.\label{eq:Ox-Neel}\end{equation}
 We should stress that the critical exponent in equation (\ref{eq:Ox-crit})
differs from the one in equations (\ref{eq:Ox-Haldane}) and (\ref{eq:Ox-Neel});
it is not possible to recover the decay behaviour at $h=1$ from the
functions found for $h>1$ or $h<1$ simply by letting $R/\xi\to0$
in the exponentials. Qualitatively, the reason is that the correlation
functions should be described by a unique scaling function ${\cal F}(r)$
of the variable $r=R/\xi$, but the asymptotic expansions in the off-critical
regime and in the critical regime are different. The former corresponds
to $r\gg1$ while the latter to $r\to0$ for any large but finite
value of $R$. A similar argument holds also for the longitudinal
spin-spin correlation function ${\cal C}_{z}(R)$ of the previous
subsection. Although possible in principle (see, for instance, \cite{wmtb1976}
for the 2D classical Ising model), the derivation and the usage of
the whole scaling functions is beyond the scope of this paper. Finally,
we notice from the equations above, as compared to those of subsec.
(\ref{sub:CzGH}), that the correlators ${\cal O}_{x}(R)$ and ${\cal C}_{z}(R)$
play a dual role above and below the transition line; when one order
parameter is vanishing, the other is not. Here we do not have an explicitly
duality relation between order and disorder lattice operators for
the spin-1 model as in the Ising case (see, however, ref. \cite{h1994}
for the XY chain). Hence, what is a nontrivial fact is to see that
also the decay laws interchange when the transition line is crossed.
\end{itemize}

\subsection{Transverse spin-spin correlation ${\cal C}_{x}(R)$}

From the analysis reported in the Appendix, we can prove a conjecture
already put forth in ref. \cite{g1989}, namely that the transverse
correlation function decays always exponentially even when one crosses
the critical line. Here we can be more precise and derive also the
power-law terms in front of the exponential\begin{equation}
{\cal C}_{x}(R)\simeq\frac{\exp(-R/\Xi)}{R^{\eta_{x}}}\;,\;\;\Xi\equiv\frac{2}{\beta+\beta'},\label{eq:abCx}\end{equation}
 where $\beta$ and $\beta'$ in the Haldane phase, along the critical
line and in the Néel phase take, respectively, the form written in
equations (\ref{eq:betaH}), (\ref{eq:betapH}), (\ref{eq:betac}),
(\ref{eq:betapc}), (\ref{eq:betaN}) and (\ref{eq:betapN}) in the
Appendix. In particular we have checked that both $\beta_{{\rm c}}$
and $\beta'_{c}$ for $h=1$ and $\gamma\ge1$ are nonzero. Hence,
despite the fact that the system is critical, the transverse correlation
function exhibits a \textit{finite} characteristic length $\Xi$.
As far as the exponent $\eta_{x}$ is concerned:

\begin{itemize}
\item \textit{Haldane phase $h<1$} ($\lambda_{2}>1$) \textit{and Néel
phase$h>1$} ($\lambda_{2}<1$): $\eta_{x}=1/2$; 
\item \textit{Critical line} $h=1$ ($\lambda_{2}=1$): $\eta_{x}=1/4$.
Despite the fact that $\Xi(h=1)<\infty$, the algebraic prefactor
is the same power-law that describes critical correlations in the
quantum Ising model.
\end{itemize}
From the values of $h$ and $\gamma$ reported in the last two columns
of table \ref{tab:n0ABhgamma} we have computed $\Xi[h(\lambda,D),\gamma(\lambda,D)]$;
for example when $\lambda=1$ we find that $\Xi$ decreases steadily
as $D$ is decreased towards larger negative values, passing from
the Haldane to the Néel phase. This behaviour is consistent with the
numerical best-fit estimates of $\Xi$ made in the next section.

\section{Comparison with DMRG results\label{sec:DMRG}}

The results of the previous section regarding the long-distance decay
of ordinary and string correlation functions are summarized in table
\ref{tab_fitt} where:\begin{equation}
f_{0}(R)=A_{0}\frac{\exp(-R/A_{1})}{R^{1/4}}\label{eq:f0}\end{equation}
 \begin{equation}
f_{1}(R)=A_{2}+A_{0}\frac{\exp(-R/A_{1})}{\sqrt{R}}\label{eq:f1}\end{equation}
 \begin{equation}
f_{2}(R)=A_{2}+A_{0}\frac{\exp(-2R/A_{1})}{R^{2}}.\label{eq:f2}\end{equation}
 Within the approximation of hidden order and for large $R$ these
asymptotic laws are to be considered exact and valid for the Haldane
and Néel phases and associated transition line as specified in table
\ref{tab_fitt}. It should be noted that $f_{1}(R)$ and $f_{2}(R)$
agree with the general form argued for the $d$-dimensional Ising
model (see, for example, \cite{c1972} and refs. therein) although
the derivation of the latter did not include the case of string correlation
functions.

The two functional forms $f_{1,2}$ now can be used to extract, for
example, the asymptotic value of string order correlation functions
computed numerically; in this sense $A_{0}$, $A_{1}$ and $A_{2}$
may be regarded as free fitting parameters. The goodness of the best-fit
procedure can be assessed by computing the reduced $\chi^{2}$: \[
\tilde{\chi}^{2}\equiv\frac{\sum\mbox{squares of differences}}{\mbox{\# of data points}-\mbox{\# of fit parameters}-1}.\]

\begin{table}

\caption{Expected asymptotic behaviour of string (${\cal O}$) and usual (${\cal C}$)
correlation functions in the Haldane and Néel phases of model (\ref{eq:HlamD})
and along the critical transition line separating them. The fitting
functions $f_{0,1,2}$ are defined in equations (\ref{eq:f0})-(\ref{eq:f2}).
Note the interchanged role of ${\cal O}_{x}(R)$ and ${\cal C}_{z}(R)$
above and below the transition line.\label{tab_fitt}}

\begin{tabular}{@{}lll}
\hline 
Phase&
C.f.&
Decay law\tabularnewline
\hline 
Haldane&
${\cal C}_{z}$&
$f_{1}$($A_{2}\equiv0$)\tabularnewline
Transition&
${\cal C}_{z}$&
$f_{0}$ ($A_{1}^{-1}=0$)\tabularnewline
Néel&
${\cal C}_{z}$&
$f_{2}$\tabularnewline
Haldane&
${\cal C}_{x}$&
$f_{1}$($A_{2}\equiv0$)\tabularnewline
Transition&
${\cal C}_{x}$&
$f_{0}$\tabularnewline
Néel&
${\cal C}_{x}$&
$f_{1}$($A_{2}\equiv0$)\tabularnewline
Haldane&
${\cal O}_{z}$&
$f_{2}$\tabularnewline
Transition&
${\cal O}_{z}$&
$f_{2}$ ($A_{1}^{-1}=0$)\tabularnewline
Néel&
${\cal O}_{z}$&
$f_{2}$\tabularnewline
Haldane&
${\cal O}_{x}$&
$f_{2}$\tabularnewline
Transition&
${\cal O}_{x}$&
$f_{0}$ ($A_{1}^{-1}=0$)\tabularnewline
Néel&
${\cal O}_{x}$&
$f_{1}$($A_{2}\equiv0$)\tabularnewline
\hline
\end{tabular}
\end{table}

Clearly one could use many other different functions to extrapolate
the correlators to $R\to\infty$. However, as recalled in the introduction,
the literature contains very few, empirical, information about the
asymptotic approach to the limit values of the string correlation
functions. Our study was motivated by this fact and so here we perform
a comparison between $f_{0}$, $f_{1}$ and $f_{2}$ by examining
their capability to fit the spin-spin and string correlations evaluated
numerically through the DMRG. Actually, following the idea of ref.
\cite{cvetal2002}, in order to take into account the PBC we employ
the left-right symmetrized expressions of equations (\ref{eq:f0})-(\ref{eq:f2})
\begin{equation}
F_{\ell}(R)\equiv\frac{f_{\ell}(R)+f_{\ell}(L-R)}{2},\;\ell=0,1,2,\label{eq:symF}\end{equation}
 at least for the correlation functions in the $z$-channel. As regards
${\cal O}_{x}$, while translational invariance implies that it depends
on the difference between the sites at the ends of the string, it
is not always guaranteed that it depends only on the distance on the
ring. In other terms the expectation value\[
\langle S_{i}^{x}{\rm e}^{{\rm i}\pi\sum_{k=i+1}^{j-1}S_{k}^{x}}S_{j}^{x}\rangle\]
 may differ from the same expression with $i$ and $j$ interchanged.
In fact, using the properties of the exponentials of spin-1 operators,
$\exp({\rm i}\pi S_{i}^{x})=\exp(-{\rm i}\pi S_{i}^{x})$ and $S_{i}^{x}\exp({\rm i}\pi S_{i}^{x})=-S_{i}^{x}$,
it can be shown that the expression above can rewritten as\[
\langle S_{j}^{x}{\rm e}^{{\rm i}\pi\sum_{k=j+1}^{i-1}S_{k}^{x}}S_{i}^{x}{\rm e}^{{\rm i}\pi S_{{\rm tot}}^{x}}\rangle\]
 where $S_{{\rm tot}}^{x}=\sum_{i=1}^{L}S_{i}^{x}$. The point is
that in general the GS of an anisotropic spin chain is not invariant
under the action of $\exp({\rm i}\pi S_{{\rm tot}}^{x})$ and so a
direct inspection is required case by case in order to decide if a
symmetrized fitting function has to be used or not.

\begin{figure}
\includegraphics[clip,width=7cm]{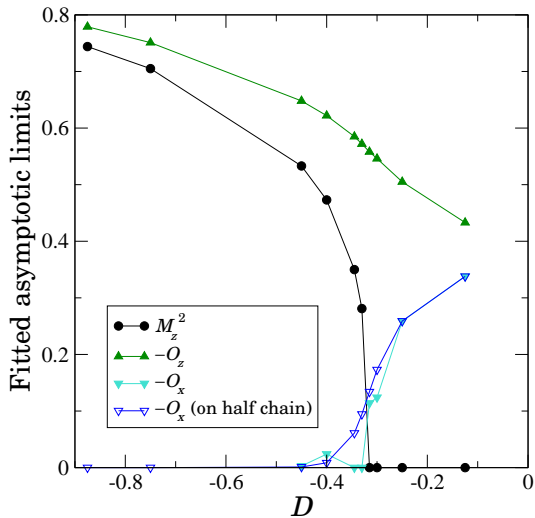}

\caption{Asymptotic values (order parameters) attained by ${\cal C}_{z}$
(dots), ${\cal O}_{z}$ (up triangles) and ${\cal O}_{x}$ (down triangles).
The limits to $R\to\infty$ correspond to the values of the best-fit
parameters $A_{2}$ for the fitting function that, case by case, gives
the smallest value of $\tilde{\chi}^{2}$. The empty triangles result
from fitting the transverse string correlation functions on half chain
(see text for explanation).\label{fig:al_vs_D}}
\end{figure}

The asymptotic limits (i.e. the values of $A_{2}$) resulting from
of a series of best-fits made on DMRG data obtained by fixing $\lambda=1$
and letting $D$ to vary across the Haldane-Néel transition from $-0.125$
to $-0.875$ are plotted in figure \ref{fig:al_vs_D}. It is seen
that the nonvanishing values of ${\cal O}_{x}$ and $M_{z}$ characterize,
respectively, the Haldane and the Néel phase. It is reasonable to
expect that the location of the critical point as the value of $D$
at which the two order parameters vanish leads to two slightly different
estimates. However, with more accurate methods the critical point
was previously found to be $D_{{\rm c}}=-0.315$ \cite{chs2003,debo2004}.

In the simulations we have fixed the total length of the chain to
be $L=100$ sites and computed the GS properties by retaining from
243 to 324 DMRG states in the sector with $S_{{\rm tot}}^{z}=0$,
which is the only good quantum number that we could use. All the functional
forms derived above are asymptotic so we cannot expect them to be
reliable for very short distances. Therefore, we have conventionally
excluded the data with $R\le5$ from the fitted points. In the Néel
phase the GS tends to become doubly degenerate in the limit $L\to\infty$;
in order to take into account this difficulty we have built the reduced
density matrix by targeting the two low-lying states rather than just
the GS. Finally we have performed three finite-system sweeps to achieve
a better accuracy. In the cases we have considered, the transverse
string correlation ${\cal O}_{x}(R)$ turned out to be symmetric with
respect to the middle of the chain except for $D=-0.75$ and $D=-0.875$.
For this reason we have repeated the fit using directly the functions
of equations (\ref{eq:f0}), (\ref{eq:f1}) and (\ref{eq:f2}) without
symmetrization selecting only the points in the first half of the
chain. The asymptotic values are essentially unaffected, with the
exception of those referring to the critical point. In general when
the results of the fit are such that $A_{1}\gg L$ (typically close
to criticality) we conclude that the exponential tail of the fitting
function is essentially saturated to unity and an algebraic fit would
produce the same result.

As far as the best-fitting functions for ${\cal C}_{z}(R)$ and ${\cal O}_{x}(R)$
are concerned, the passage from the type $f_{1}$ to the type $f_{2}$
going through $f_{0}$ at the critical point, as in table \ref{tab_fitt},
actually takes place gradually in the interval $D\in(-0.345,-0.315)$,
the worst values of $\tilde{\chi}^{2}$ being of order $10^{-5}$.
The best choice to fit the transverse spin-spin correlation function
${\cal C}_{x}(R)$, instead, follows the prediction of table \ref{tab_fitt}
($F_{1}$ except at the critical line, where it becomes $F_{0}$)
with a deviation $\tilde{\chi}^{2}<10^{-8}$. Finally, the longitudinal
string correlator ${\cal O}_{z}(R)$ is very well fitted by $F_{2}$,
in agreement with table \ref{tab_fitt}, with $\tilde{\chi}^{2}\sim10^{-9}$
or better.

It is also important to check quantitatively the goodness of the Hartree-Fock
approximation. The decoupling parameter in the fermionic version are
$n_{0}=\langle n_{j=0}\rangle$, $A=\langle c_{1}^{\dagger}c_{0}\rangle$
and $B=\langle c_{1}c_{0}\rangle$ where we have selected a reference
site {}``0'' invoking translational invariance. In the original
spin-1 formalism it can be checked directly that\begin{equation}
n_{0}=\langle(S_{0}^{z})^{2}\rangle\;,\;\; A=\frac{1}{2}\langle S_{1}^{z}\left(S_{0}^{+}S_{1}^{-}+S_{1}^{+}S_{0}^{-}\right)S_{0}^{z}\rangle\label{eq:nA}\end{equation}

The operator $c_{1}c_{0}$ destroys a couple of fermions in adjacent
sites; in the spin language they could be $\uparrow\downarrow$ or
$\downarrow\uparrow$ depending on the surrounding sites in order
to respect the AFM order. Let us express the GS in the form $\vert{\rm GS}\rangle=\alpha\vert\Uparrow\rangle+\beta\vert\Downarrow\rangle$,
where $\vert\Uparrow\rangle$ denotes a linear combination of states
in which the first nonzero spin along-$z$ is directed upward and
$\vert\Downarrow\rangle$ the same state with all the spin reversed.
Only one of the terms in$\left(S_{0}^{+}S_{1}^{-}+S_{1}^{+}S_{0}^{-}\right)$
will act on $\vert\Uparrow\rangle$ respecting the AFM order and the
other term will thereby act on $\vert\Downarrow\rangle$. When the
scalar product with $\langle{\rm GS\vert}$ is taken, the states from
$\vert\Uparrow\rangle$ will not mix with those from $\vert\Downarrow\rangle$.
Therefore we try with the expression\begin{equation}
B=-\frac{1}{2}\langle\left(S_{0}^{-}S_{1}^{+}+S_{0}^{+}S_{1}^{-}\right)S_{0}^{z}S_{1}^{z}\rangle.\label{eq:B}\end{equation}

In table \ref{tab:n0AB-DMRG-sc} we report the values of the decoupling
parameters for a set of points in the Haldane and Néel phases, comparing
the DMRG values with the numerical solution of the self-consistent
equations using 100 iterations from different choices of initial conditions.
Having the DMRG estimates for $n_{0}$ and $A$ we may also produce
a {}``hybrid'' estimate of the critical point by setting $h=1$
in equation (\ref{eq:hgamma}) and then solving for $\tilde{D}_{{\rm c}}(\lambda)=2[\lambda(n_{0,{\rm DMRG}}-A_{{\rm DMRG}})-1]$.
With $\lambda=1$ we find for example $\tilde{D}_{{\rm c}}=-0.254$,
that compares slightly better than the fully-self-consistent value
($D_{{\rm c}}=-0.214$) to the accepted numerical one $D_{{\rm c}}\cong-0.315$.

Apart from the value $n_{0}$, which quantifies the number of spins
with nonzero projection along $z$, we expect that the goodness of
the mapping used in this work is higher when the hidden order is larger.
Therefore, as a final check, we have repeated the passages of Sect.
\ref{sec:Mapping} (see \cite{tdMario} for details) including also
a biquadratic term $\frac{1}{3}\sum_{i}(\vec{S}_{i}\cdot\vec{S}_{i+1})^{2}$
in the spin-1 Hamiltonian (\ref{eq:HlamD}). For $\lambda=1$ and
$D=0$ the ground state of the model can be found exactly \cite{aklt1988,kt1992}
using the valence-bond picture: each spin-1 is viewed as the triplet
sector of a pair of spin-1/2 particles and the ground state is constructed
by creating a sequence of singlets between adjacent sites. In this
case the string correlation functions can be computed exactly and
it turns out that ${\cal O}_{x,z}(R)=-4/9$ independent of $R$. At
the isotropic point with biquadratic term the self-consistent equations
are solved by $n_{0}=2/3$, $A=B=-2/9$ and the effective parameters
of the XY model become $h=3/5$ and $\gamma=4/5$. From equation (\ref{eq:lambda12})
we find $\lambda_{1}=\lambda_{2}=3$ so that in equation (\ref{eq:Oz-vs-sigmaz})
we have just $\langle\sigma_{j}^{z}\rangle=1/3$ and ${\cal O}_{z}=-4/9$.
Interestingly enough, even if the XY model does not have an explicit
rotational symmetry as the original spin-1 Hamiltonian, by inserting
these values of $h$ and $\gamma$ into the constant part of equation
(\ref{eq:Ox-Haldane}) we find again ${\cal O}_{x}=-4/9$. This accordance
can be taken as a positive check of our approach.

\begin{table}

\caption{DMRG $(L=100)$ versus self-consistent (s-c) estimates of the three
decoupling parameters $n_{0}$, $A$ and \textbf{$B$} of equations
(\ref{eq:nA}) and (\ref{eq:B}). It must be kept in mind that the
continuum versions of the self-consistent equations neglect some $O(L^{-1})$
terms coming from isolated contributions at wavenumber 0 or $\pi$.\label{tab:n0AB-DMRG-sc}}

\begin{tabular}{llllllll}
\hline 
$\lambda$&
$D$&
$n_{{\rm 0,DMRG}}$&
$A_{{\rm DMRG}}$&
$B_{{\rm DMRG}}$&
$n_{0,{\rm s-c}}$&
$A_{{\rm s-c}}$&
$B_{{\rm s-c}}$\tabularnewline
\hline 
$1$&
$0$&
$0.667$&
$-0.166$&
$-0.30080\pm0.00005$&
$0.709$&
$-0.158$&
$-0.253$\tabularnewline
$1$&
$-0.125$&
$0.702$&
$-0.151$&
$-0.28908\pm0.00005$&
$0.745$&
$-0.137$&
$-0.246$\tabularnewline
$1$&
$-10$&
$0.996$&
$-0.000324$&
$-0.0442$&
$0.996$&
$-0.000317$&
$-0.0433$\tabularnewline
$5$&
$-0.125$&
$0.991$&
$-0.000860$&
$-0.0654$&
$0.991$&
$-0.000853$&
$-0.0649$\tabularnewline
\hline
\end{tabular}

\end{table}

\section{Conclusions\label{sec:Conc}}

In this paper we have reconsidered and extended the approach of ref.
\cite{g1989} to the GS properties of spin-1 anisotropic quantum chains.
We have included a single-ion term in the Hamiltonian and, moreover,
we have analyzed explicitly how the spin-1 correlation functions are
written in the spinless fermions language and then in the framework
of the XY model in a transverse field for effective spin-1/2 degrees
of freedom. In particular, we have focused on the decay laws of the
string correlators towards their asymptotic values which apparently
were missing in the literature.

The decay laws of string and spin-spin correlation functions (in the
longitudinal channel) are all related to the generating function $c({\rm e}^{{\rm i}k})$
of equation (\ref{eq:GR}) and the determinants of the Toeplitz matrices
derived from it. The asymptotic behaviour of the transverse correlation
function ${\cal C}_{x}(R)$, instead, originates from a product of
two Toeplitz determinants (see the Appendix, in particular eq. (\ref{eq:diagM1})).
The leading terms in the regime $R\gg1$ for the various correlators
are discussed in Sect. \ref{sec:adl} and summarized in table \ref{tab_fitt}.
In brief the most interesting points unveiled by the approach used
here are:

\begin{itemize}
\item The nonvanishing string-order parameters of the spin-1 model (\ref{eq:HlamD})
are simply interpreted as the magnetization along $x$ and $z$ in
the XY chain with transverse field (eqs. (\ref{eq:Oz-vs-sigmaz})
and (\ref{eq:Ox-Haldane})). 
\item There exists dual behaviour of ${\cal O}_{x}(R)$ and ${\cal C}_{z}(R)$
above and below the transition, both for the asymptotic order parameters
and for the decay functional forms. 
\item The explicit calculation of ${\cal C}_{x}(R)$ allows us to prove
an unusual feature in statistical mechanics, already conjectured by
G\'{o}mez-Santos \cite{g1989}: the spin-spin transverse correlation
function exhibits always a finite characteristic length $\Xi$ (eq.
(\ref{eq:abCx})) even when the system becomes critical.
\end{itemize}
The analytical results are supported by comparison with a numerical
(DMRG) study of the model, especially for the correlations ${\cal C}_{x}$
and ${\cal O}_{z}$. A more detailed comparison between the analytical
and the numerical estimates should take into account: i) finite-size
effects due to a finite total length $L$ while in Sect. \ref{sec:Mapping}
we passed readily to the thermodynamic limit; ii) corrections for
finite distance $R$ beyond the dominant ones. Although in principle
they can be computed systematically, in this paper we have limited
ourselves to the leading terms in order to derive analytical expressions
with the smallest possible number of fitting parameters.

\section*{Acknowledgments}

This work was partially supported by the Italian MiUR through the
PRIN grant n. 2007JHLPEZ. M.R. acknowledges support from the EU (SCALA).

\appendix

\section*{Appendix: Toeplitz formulation of ${\cal C}_{x}(R)$}

The fermionic version of equation (\ref{eq:S+S--sigma}) reads\[
\frac{1}{2}\langle S_{j}^{+}S_{j+R}^{-}+S_{j}^{-}S_{j+R}^{+}\rangle=\langle A_{j}\prod_{k<j}\left(1-2n_{k}\right)\prod_{k=j+1}^{j+R-1}(1-n_{k})\prod_{k<j+R}\left(1-2n_{k}\right)A_{j+R}\rangle\]
 \[
=\langle B_{j}\left(\prod_{k=j+1}^{j+R-1}c_{k}c_{k}^{\dagger}\right)A_{j+R}\rangle=\langle c_{j}^{\dagger}\left(\prod_{k=j+1}^{j+R-1}c_{k}c_{k}^{\dagger}\right)c_{j+R}^{\dagger}\rangle\]
 \[
+\langle c_{j}^{\dagger}\left(\prod_{k=j+1}^{j+R-1}c_{k}c_{k}^{\dagger}\right)c_{j+R}\rangle-\langle c_{j}\left(\prod_{k=j+1}^{j+R-1}c_{k}c_{k}^{\dagger}\right)c_{j+R}^{\dagger}\rangle-\langle c_{j}\left(\prod_{k=j+1}^{j+R-1}c_{k}c_{k}^{\dagger}\right)c_{j+R}\rangle\]
 By observing that\[
\langle c_{j}^{\dagger}\left(\prod_{k=j+1}^{j+R-1}c_{k}c_{k}^{\dagger}\right)c_{j+R}^{\dagger}\rangle=\left\langle \left(c_{j}^{\dagger}\left(\prod_{k=j+1}^{j+R-1}c_{k}c_{k}^{\dagger}\right)c_{j+R}^{\dagger}\right)^{\dagger}\right\rangle =-\langle c_{j}\left(\prod_{k=j+1}^{j+R-1}c_{k}c_{k}^{\dagger}\right)c_{j+R}\rangle\]
 \[
\langle c_{j}^{\dagger}\left(\prod_{k=j+1}^{j+R-1}c_{k}c_{k}^{\dagger}\right)c_{j+R}\rangle=\left\langle \left(c_{j}^{\dagger}\left(\prod_{k=j+1}^{j+R-1}c_{k}c_{k}^{\dagger}\right)c_{j+R}\right)^{\dagger}\right\rangle =-\langle c_{j}\left(\prod_{k=j+1}^{j+R-1}c_{k}c_{k}^{\dagger}\right)c_{j+R}^{\dagger}\rangle\]
 we can write ${\cal C}_{x}(R)=(-1)^{R}\langle S_{j}^{+}S_{j+R}^{-}+S_{j}^{-}S_{j+R}^{+}\rangle/4$
(using translational and U(1) rotational invariance about $z$) as\begin{equation}
{\cal C}_{x}(R)=-\sqrt{\det{\bf M}_{1}}-\sqrt{\det{\bf M}_{2}}\label{eq:Cx-sqrtdet}\end{equation}
 where the two terms come, respectively, from the Pfaffians\[
\left.\begin{array}{cccccccccc}
{\rm Pf}|{\rm i}F_{1} & {\rm i}F_{2} & \cdots & {\rm i}F_{R-2} & {\rm i}F_{R-1} & -H_{-1} & -H_{-2} & \cdots & -H_{-R+1} & -H_{-R}\\
 & {\rm i}F_{1} & \cdots & {\rm i}F_{R-3} & {\rm i}F_{R-2} & -H_{0} & -H_{-1} & \cdots & -H_{-R+2} & {-H}_{-R+1}\\
 &  & \ddots & \vdots & \vdots & \vdots & \vdots & \vdots & \vdots & \vdots\\
 &  &  & {\rm i}F_{1} & {\rm i}F_{2} & -H_{R-4} & -H_{R-5} & \cdots & -H_{-2} & {-H}_{-3}\\
 &  &  &  & {\rm i}F_{1} & -H_{R-3} & -H_{R-4} & \cdots & -H_{-1} & -H_{-2}\\
 &  &  &  &  & -H_{R-2} & -H_{R-3} & \cdots & -H_{0} & -H_{-1}\\
 &  &  &  &  &  & -{\rm i}F_{1} & \cdots & -{\rm i}F_{R-2} & -{\rm i}F_{R-1}\\
 &  &  &  &  &  &  & \ddots & \vdots & \vdots\\
 &  &  &  &  &  &  &  & -{\rm i}F_{1} & -{\rm i}F_{2}\\
 &  &  &  &  &  &  &  &  & -{\rm i}F_{1}\end{array}\right|\]
 and\[
\left.\begin{array}{cccccccccc}
{\rm Pf|}{\rm i}F_{1} & {\rm i}F_{2} & \cdots & {\rm i}F_{R-2} & {\rm i}F_{R-1} & {-H}_{-1} & {-H}_{-2} & \cdots & {-H}_{-R+1} & {\rm i}F_{R}\\
 & {\rm i}F_{1} & \cdots & {\rm i}F_{R-3} & {\rm i}F_{R-2} & -H_{0} & -H_{-1} & \cdots & {-H}_{-R+2} & {\rm i}F_{R-1}\\
 &  & \ddots & \vdots & \vdots & \vdots & \vdots & \vdots & \vdots & \vdots\\
 &  &  & {\rm i}F_{1} & {\rm i}F_{2} & -H_{R-4} & -H_{R-5} & \cdots & -H_{-2} & {\rm i}F_{3}\\
 &  &  &  & {\rm i}F_{1} & -H_{R-3} & -H_{R-4} & \cdots & -H_{-1} & {\rm i}F_{2}\\
 &  &  &  &  & -H_{R-2} & -H_{R-3} & \cdots & -H_{0} & {\rm i}F_{1}\\
 &  &  &  &  &  & -{\rm i}F_{1} & \cdots & -{\rm i}F_{R-2} & H_{R-1}\\
 &  &  &  &  &  &  & \ddots & \vdots & \vdots\\
 &  &  &  &  &  &  &  & -{\rm i}F_{1} & H_{2}\\
 &  &  &  &  &  &  &  &  & H_{1}\end{array}\right|\]
 with\[
F_{l-j}\equiv{\rm i}\langle c_{j}c_{l}\rangle=-{\rm i}\langle c_{j}^{\dagger}c_{l}^{\dagger}\rangle=\frac{1}{2\pi}\int_{0}^{2\pi}{\rm d}k{\rm e}^{-{\rm i}k(l-j)}f\left({\rm e}^{{\rm i}k}\right)\]
 \[
H_{l-j}\equiv\langle c_{j}c_{l}^{\dagger}\rangle=\frac{1}{2\pi}\int_{0}^{2\pi}{\rm d}k{\rm e}^{-{\rm i}k(l-j)}h\left({\rm e}^{{\rm i}k}\right){\rm e}^{-{\rm i}k}\]
 \[
f\left({\rm e}^{{\rm i}k}\right)\equiv\frac{\gamma\sin k}{2\sqrt{(\cos k-h)^{2}+\gamma^{2}\sin^{2}k}}\;,\;\; h\left({\rm e}^{{\rm i}k}\right)\equiv\frac{{\rm e}^{{\rm i}k}}{2}\left(1+\frac{\cos k-h}{\sqrt{(\cos k-h)^{2}+\gamma^{2}\sin^{2}k}}\right).\]
 It is useful to note that $F_{l-j}=-F_{j-l}$ and $H_{l-j}=H_{j-l}$.
According to usual conventions, the Toeplitz matrix\[
{\bf M}_{1}=\left(\begin{array}{cc}
-{\rm i}{\bf F} & -{\bf H}\\
{\bf H^{{\rm T}}} & +{\rm i}{\bf F}\end{array}\right)={\bf M}_{1}[\phi]\;,\;\;{\bf M}_{2}={\bf M}_{1}+{\bf M}_{0}\]
 in equation (\ref{eq:Cx-sqrtdet}) is generated by matrix-valued
symbol (analytically continued to the unit circle)\[
\phi(z)=\left(\begin{array}{cc}
-{\rm i}f(z) & -h(z)\\
h(z^{-1}) & {\rm i}f(z^{-1})\end{array}\right)\]
 while ${\rm {\bf M}}_{2}={\rm {\bf M}}_{1}-{\rm {\bf M}}_{0}$ with\[
{\bf M}_{0}=\left(\begin{array}{ccccccc}
0 & \cdots & 0 & 0 & \dots & 0 & -H_{-R}-{\rm i}F_{R}\\
\vdots & \vdots & \vdots & \vdots & \vdots & \vdots & \vdots\\
0 & \dots & 0 & 0 & \dots & 0 & -H_{-1}-{\rm i}F_{1}\\
0 & \cdots & 0 & 0 & \dots & 0 & -H_{R-1}-{\rm i}F_{R-1}\\
\vdots & \vdots & \vdots & \vdots & \vdots & \vdots & \vdots\\
0 & \dots & 0 & 0 & \dots & 0 & -H_{1}-{\rm i}F_{1}\\
H_{-R}+{\rm i}F_{R} & \cdots & H_{-1}+{\rm i}F_{1} & H_{R-1}+{\rm i}F_{R-1} & \dots & H_{-1}+{\rm i}F_{1} & 0\end{array}\right).\]

Now, since the trace norm of ${\rm {\bf M}}_{0}$ is vanishing, when
$R\to\infty$ we get $\det{\bf M}_{1}=\det{\bf M}_{2}$ so that so
that ${\cal C}_{x}(R)\simeq-2\sqrt{\det{\bf M}_{1}}$. In order to
compute the determinant, we need to check the possible zeroes of\[
\det\phi(z)=\frac{1}{2}\left(1+\mbox{sign}(z)\frac{z^{2}+1-2hz}{\sqrt{\left(z^{2}+1-2hz\right)^{2}-\gamma^{2}\left(z^{2}-1\right)^{2}}}\right);\]
 for instance, when $z=\pm1$, $\det\phi(\pm1)=\frac{1}{2}\left(1+\mbox{sign}(\pm1)\mbox{sign}\left(1\mp h\right)\right)$
so that for $h<1$ the symbol is singular at $z=-1$, while for $h>1$
it is singular at $z=+1$. Unfortunately, as discussed also in ref.
\cite{l2007}, known results for matrix-valued symbol do not cover
the case of singular symbols with vanishing determinant. Hence, the
strategy is to factorize the determinant of ${\rm {\bf M}}_{1}$ as
a product of determinants of matrices generated by scalar-valued symbols.
Fortunately, in this case this task is accomplished by transforming
${\rm {\bf M}}_{1}$ through the matrix\[
{\bf U}=\left(\begin{array}{cc}
{\bf 1} & {\bf {\rm i}}{\rm {\bf F}}^{-1}{\rm {\bf H}}\\
{\bf 0} & {\bf {\bf 1}}\end{array}\right)\]
 so that\begin{equation}
{\rm {\bf U}}^{{\rm T}}{\rm {\bf M}}_{1}{\rm {\bf U}}=\left(\begin{array}{cc}
-{\rm i}{\rm {\bf F}} & {\rm {\bf 0}}\\
{\rm {\bf 0}} & {\rm i}{\rm {\bf F}}+{\rm i}{\rm {\bf H}}^{{\rm T}}{\rm {\bf F}}^{-1}{\rm {\bf H}}\end{array}\right).\label{eq:diagM1}\end{equation}
Now, we first use a theorem by Widom and Silbermann (see, for instance,
\cite{bg1991,bs1998}) according to which ${\rm {\bf F}}^{-1}$ is
a Toeplitz matrix generated by $f^{-1}$ (in the present case this
result holds for even $R$) and then express the product ${\rm {\bf H}}^{{\rm T}}{\rm {\bf F}}^{-1}{\rm {\bf H}}$
as another Toeplitz matrix generated by the symbol ${\rm i}h(z^{-1})f^{-1}(z)h(z)$.
The last identification can be done by using repeatedly a theorem
by Brown and Halmos \cite{bh1963}:

\begin{quotation}
\textit{$T(\varphi)T(\psi)$ is a Toeplitz operator iff either $\varphi^{*}(z)$
or $\psi(z)$ are analytic functions; if the latter condition is satisfied
then $T(\varphi)T(\psi)=T(\varphi\psi)$}
\end{quotation}
where $T(\varphi)$ denotes the Toeplitz matrix generated by the function
$\varphi$.

Let us start by computing $\det(-{\rm i}{\rm {\bf F}})$:

\begin{itemize}
\item \textit{Haldane phase $h<1$} ($\lambda_{2}>1$). From the analytic
continuation to the unit circle\begin{equation}
f(z)=-{\rm i}\frac{\gamma}{2\left(1-\gamma\right)}\mbox{sign}(z)\frac{\left(1-z\right)\left(1+z\right)}{\sqrt{\left(z-\lambda_{1}\right)\left(z-\lambda_{2}\right)\left(1-z\lambda_{1}\right)\left(1-z\lambda_{2}\right)}}\label{eq:fz}\end{equation}
 we see that $f$ vanishes at $z=\pm1$ and is singular at $z=\lambda_{1,2}^{-1}$.
This case is covered by the Fisher-Hartwig conjecture (see refs. \cite{bg1991,bs1998,fa2005,l2007})
and the asymptotic behaviour turns out to be $\det(-{\rm i}{\rm {\bf F}})\sim E_{{\rm H}}R^{-1}\exp(-\beta_{{\rm H}}R)$
with\begin{equation}
\beta_{{\rm H}}=\frac{7}{4}\ln[\vert\lambda_{1}\vert\lambda_{2}]-\ln\frac{\gamma}{2(\gamma-1)}.\label{eq:betaH}\end{equation}
In general the exponent of the power-law prefactor is given by $\sum_{r}\alpha_{r}^{2}-\beta_{r}^{2}$
where the index $r$ runs over all zero and singular points $z_{r}$
of $f(z)$ while the numbers $\alpha_{r}$ and $\beta_{r}$ are defined
through the factorization of the function in the following form\[
-{\rm i}f(z)=\tau(z)\prod_{r}\left(1-\frac{z}{z_{r}}\right)^{\alpha_{r}+\beta_{r}}\left(1-\frac{z_{r}}{z}\right)^{\alpha_{r}-\beta_{r}},\]
the residual function $\tau(z)$ satisfying the conditions of Szegö's
theorem. In this specific case, the $1/R$ prefactor is due to the
combination of the exponents reported in table \ref{tab:zabH}. Finally,
the constant prefactors $E$ can also be written down explicitly in
the framework of the Fisher-Hartwig conjecture, although the result
will not be given here for the sake of brevity and because they will
be treated as free fitting parameters.
\item \textit{Critical line} $h=1$ ($\lambda_{2}=1)$. There is only one
zero at $z=-1$ and one singularity at $z=1/\lambda_{1}$. The exponents
associated with these two points are the same as in the Haldane phase;
in this case the power of $R$ receives contributions only from these
two points and becomes $[(1/2)^{2}-(3/4)^{2}+(-1/4)^{2}-(-1/2)^{2}]=-1/2$,
instead of $-1$. However, the characteristic inverse scale in the
exponential is nonvanishing even at the critical point\begin{equation}
\beta_{{\rm c}}=\frac{7}{4}\ln\vert\lambda_{1}\vert-\ln\frac{\gamma}{2(\gamma-1)}.\label{eq:betac}\end{equation}

\item \textit{Néel phase $h>1$} ($\lambda_{2}<1$). The zeroes remain at
$z=\pm1$ while the singularities now are at $z=\lambda_{1}^{-1}$
and $z=\lambda_{2}$. Therefore, we proceed along the same line followed
for the Haldane phase, just by replacing $\lambda_{2}\leftrightarrow1/\lambda_{2}$.
In particular, we find the same numbers $\alpha_{r}$ and $\beta_{r}$
as for the case $h<1$ and thus the asymptotic behaviour remains of
the form $\det(-{\rm i}{\rm {\bf F}})\sim E_{{\rm N}}R^{-1}\exp(-\beta_{{\rm N}}R)$
with\begin{equation}
\beta_{{\rm N}}=\frac{7}{4}\ln\frac{\vert\lambda_{1}\vert}{\lambda_{2}}-\ln\frac{\gamma}{2(\gamma-1)}.\label{eq:betaN}\end{equation}

\end{itemize}
\begin{table}

\caption{Values of $z_{r},\alpha_{r},\beta_{r}$ for the function $f(z)$
(eq. (\ref{eq:fz})) in the Haldane and Néel phases.\label{tab:zabH} }

\begin{tabular}{lll}
\hline 
$z_{r}$&
$\alpha_{r}$&
$\beta_{r}$\tabularnewline
\hline 
$-1$&
$1/2$&
$3/4$\tabularnewline
$+1$&
$1/2$&
$3/4$\tabularnewline
$\lambda_{1}^{-1}$&
$-1/4$&
$-1/2$\tabularnewline
$\lambda_{2}^{-1}$&
$-1/4$&
$-1/2$\tabularnewline
\hline
\end{tabular}
\end{table}

Let us now pass to $\det{\rm {\bf G}}$, with ${\rm {\bf {\rm {\bf G}}={\rm i}{\rm {\bf F}}+{\rm i}{\rm {\bf H}}^{{\rm T}}{\rm {\bf F}}^{-1}{\rm {\bf H}}}}$,
generated by the symbol\[
g(z)={\rm i}h(z^{-1})f^{-1}(z)h(z)+{\rm i}f(z)=-\frac{1}{\gamma}\frac{1}{z^{2}-1}\left[z^{2}-2hz+1+z\sqrt{\left(z+z^{-1}-2h\right)^{2}-\gamma^{2}\left(z-z^{-1}\right)^{2}}\right]\]
 (analytically continued to the unit circle).

\begin{itemize}
\item \textit{Haldane phase} $h<1$ ($\lambda_{2}>1$). The Fisher-Hartwig
conjecture now can be applied, thanks to the following factorization\begin{equation}
g(z)=\tau(z)\left(1-z\right)^{\alpha_{1}+\beta_{1}}\left(1-z^{-1}\right)^{\alpha_{1}-\beta_{1}}\left(1+z\right)^{\alpha_{2}+\beta_{2}}\left(1+z^{-1}\right)^{\alpha_{2}-\beta_{2}}\label{eq:gz}\end{equation}
 with $\alpha_{1,2}$ and $\beta_{1,2}$ as in table \ref{tab:zabg}
and with\[
\tau(z)=\frac{1}{\gamma}\frac{1}{\left(1+z\right)^{2}}\left[z^{2}-2hz+1+z\sqrt{\left(z+z^{-1}-2h\right)^{2}-\gamma^{2}\left(z-z^{-1}\right)^{2}}\right]\]
 satisfying Szegö's theorem. Consequently, the asymptotic behaviour
is purely exponential: $\det{\rm {\bf G}}\sim E'_{{\rm H}}\exp(-\beta'_{{\rm H}}R)$
where\begin{equation}
\beta'_{{\rm H}}=-\frac{1}{2\pi}\int_{0}^{2\pi}{\rm d}k\ln\left\vert \cos k-h+\sqrt{\left(\cos k-h\right)^{2}+\left(\gamma\sin k\right)^{2}}\right\vert +\ln\frac{\gamma}{2}.\label{eq:betapH}\end{equation}

\item \textit{Critical line $h=1$} $(\lambda_{2}=1$). There are no singularities
and a simple zero at $z=-1$, with exponents $\alpha$ and $\beta$
as in the first row of table \ref{tab:zabg}. Therefore, the net power
of $R$ in the algebraic prefactor vanishes and the decay is purely
exponential with\begin{equation}
\beta'_{{\rm c}}=-\frac{1}{2\pi}\int_{0}^{2\pi}{\rm d}k\ln\left\vert \cos k-1+\sqrt{\left(\cos k-1\right)^{2}+\left(\gamma\sin k\right)^{2}}\right\vert +\ln\frac{\gamma}{2}.\label{eq:betapc}\end{equation}
 As a function of $\gamma$, $\beta'_{{\rm c}}$ is decreasing for
$\gamma>1$ but does not vanish.
\item \textit{Néel phase $h>1$} $(\lambda_{2}<1$). With respect to the
Haldane phase, the function $\tau(z)$ changes to\[
\tau(z)=\frac{1}{\gamma}\frac{1}{\left(1+z\right)^{2}\left(1-z\right)^{2}}\left[z^{2}-2hz+1+z\sqrt{\left(z+z^{-1}-2h\right)^{2}-\gamma^{2}\left(z-z^{-1}\right)^{2}}\right]\]
 while the exponents $\alpha_{1,2}$ and $\beta_{1,2}$ are reported
in the fourth and fifth column of table \ref{tab:zabg}. Again, there
is no algebraic prefactor and the constant of the exponential decay,
$\det{\rm {\bf G}}\sim E'_{{\rm N}}\exp(-\beta'_{{\rm N}}R)$, reads\begin{equation}
\beta'_{{\rm N}}=-\frac{1}{2\pi}\int_{0}^{2\pi}{\rm {\rm d}}k\ln\left\vert \cos k-h+\sqrt{\left(\cos k-h\right)^{2}+\left(\gamma\sin k\right)^{2}}\right\vert +\ln\frac{\gamma}{2}=\beta'_{{\rm H}}(\gamma,h).\label{eq:betapN}\end{equation}

\end{itemize}
\begin{table}

\caption{Values of $z_{r},\alpha_{r},\beta_{r}$ for the function $g(z)$
in the Haldane and Néel phases.\label{tab:zabg}}

\begin{tabular}{lllll}
\hline 
$z_{r}$&
$\alpha_{r}$ (H)&
$\beta_{r}$ (H)&
$\alpha_{r}$ (N)&
$\beta_{r}$ (N)\tabularnewline
\hline 
$-1$&
$1/2$&
$1/2$&
$1/2$&
$1/2$\tabularnewline
$+1$&
$-1/2$&
$-1/2$&
$1/2$&
$1/2$\tabularnewline
\hline
\end{tabular}
\end{table}

\bibliographystyle{unsrt} \bibliographystyle{unsrt}
\bibliography{abpsc_JPA}

\end{document}